\documentclass{article}
\usepackage{latexsym, amsmath, amssymb, mathrsfs,
  graphics, color, multicol, epsfig}
\newcommand{\vectr}{\mathbf{r}}
\newcommand{\SAW}{\mathrm{SAW}}

\title{Non-intersection exponents of fully packed trails on the square 
lattice.}

\author{Yacine Ikhlef${}^{1,2}$, Jesper Jacobsen${}^{1,2}$ and Hubert
Saleur${}^
{2,3}$ \\[2.0mm]
${}^1$ LPTMS, Universit\'e Paris-Sud, B\^atiment 100, \\
Orsay, 91405, France \\
${}^2$ Service de Physique Th\'eorique, CEA Saclay, \\
Gif Sur Yvette, 91191, France \\
${}^3$ Department of Physics and Astronomy,
University of Southern California, \\
Los Angeles, CA 90089, USA}

\begin{document}

\maketitle

\begin{abstract}
  Fully packed trails on the square lattice are known to be described, 
  in the long distance limit, by a collection of free non compact bosons 
  and symplectic fermions, and thus exhibit some 
  properties reminiscent of Brownian motion, like vanishing fuseau 
  exponents. We investigate in this 
  paper the situation for  their non-intersection exponents. Our 
  approach is purely numerical, and based both on transfer matrix and 
  Monte Carlo calculations. We find some evidence for non-intersection 
  exponents given by CFT formulas similar  to 
  the Brownian case, albeit slightly different in their details.
\end{abstract}

\section{Introduction}

The study of conformally invariant random curves embedded in two-dimensional
space has a long history in theoretical physics and probability theory.
Recently the subject has gained new momentum due to the advent of a set of
rigorous methods known under the name of Stochastic Loewner Evolution (SLE).

Two classes of random curves have been especially well studied. In the first
class one finds self-avoiding and mutually avoiding curves, which come in
several variants depending on their precise microscopic definition, and in
particular whether they occupy a finite fraction of space in the continuum
limit. These curves admit a height representation (roughly speaking, using
the curves as level lines of the height) and as such are amenable to the
use of Coulomb Gas (CG) techniques. Also, the corresponding (twisted) vertex
models can be studied as quantum spin chains, {\it i.e.}, 
using Bethe Ansatz (BA).
As a result, the bulk critical behaviour of self-avoiding curves
is very well understood, whereas important questions regarding their surface
behaviour still remain open \cite{JS06}. Many of the exact results obtained
by the CG route have recently been confirmed, or extended, by the SLE
approach.

The second example of random curves is that of Brownian motion, which is
best studied directly in the continuum limit. Profound links exist between
this and the foregoing case, \textit{via} the use of SLE \cite{LSW01} or quantum
gravity \cite{Duplantier98} methods. To give but one example, the
external hull of the Brownian  curve turns out to be in the universality class of
the
(standard, dilute) self-avoiding walk \cite{Duplantier98}. On a microscopic
level, however, there are two crucial differences. First, a Brownian 
motion can
intersect itself, making the application of height mappings and CG techniques
break down. Second, it can pass through the same site an infinite number
of times, making it more difficult to use the 
BA approach~\footnote{We note that there has been some important 
progress recently \cite{Korchemsky} in the study of Bethe Ansatz 
for statistical mechanics systems with an infinite number of degrees
of freedom per site in relation with quantum spin chains on non 
compact groups.}.

As a result, up to this day, the 
non-intersection exponents for Brownian motion, which were identified 
by Duplantier and Kwon \cite{DK88,Duplantier98} twenty years ago, and 
whose values were rigorously established recently \cite{LSW01},
cannot be derived using 
the BA or variants of the CG method applied 
directly to a suitable discrete  lattice model in flat space.

The present work started with the recent discovery of 
lattice models that bear a close resemblance to Brownian 
motion, yet are tractable by the usual techniques. These models 
are what is sometimes called {\sl dense trails}, and  can be obtained 
by allowing self-intersections
as well as  mutual intersections in a dense loop gas, while still requiring 
 that  the curves pass through each lattice link (resp.\ site)
once (resp.\ twice) at the most. The introduction of such 
intersections is well known not to change the universality class in 
the dilute case, but---surprisingly maybe---does affect it profoundly in 
the dense case (for an early discussion  see \cite{Cao}). 
It was in particular argued in  \cite{jrs03} that for 
such a model with a loop fugacity $n<2$, all the fuseau exponents 
vanish exactly, and the corresponding correlation functions are 
described by  non compact bosonic fields, exactly like in the pure 
Brownian case. This was checked analytically and numerically in 
the case of   fully packed trails (also called 
{\em Brauer loop model}) \cite{nienhuis98,jrs03}.
 
It is of course tempting to ask how close fully packed trails and Brownian 
motion actually are, and a natural route to investigate this question 
is to study the non-intersection exponents. 

Fully packed trails, or Brauer loop models, depend among other things on the 
fugacity $n$ of loops. We will concentrate here on the case $n=0$ 
(hence, of a single trail), but notice that the case 
$n=1$  has
recently been studied numerically by Kager and Nienhuis \cite{KN06}.
These authors found that the hull distribution was compatible with that
of a reflected Brownian motion. In our work, we define a
scale-invariant ``escape path'' $\Gamma$ within a Brauer loop model
with fugacity $n=0$. Our main findings are that while the
fractal dimension of $\Gamma$ agrees (marginally) 
with that of the self-avoiding walk, the non-intersection
exponents are definitely different from those of \cite{DK88,LSW01}, 
though might well be given exactly by a related formula.

The paper is organized as follows. In section~\ref{sec:loop-model} we
define precisely the model of fully packed trails to be studied, and
in section~\ref{sec:def-gamma} we introduce the escape path $\Gamma$
and discuss how it can be made scale invariant. A qualitative phase
diagram is established in section~\ref{sec:phase-diagram}. Finally,
the non-intersection exponents and the scaling properties of $\Gamma$
are measured using transfer matrix (section~\ref{sec:transfer-matrix})
and Monte-Carlo (section~\ref{sec:MC}) techniques. We give our
conclusions in section~\ref{sec:conclusion}.

\section{Brauer loop model on the square lattice}
\label{sec:loop-model}

\begin{figure}
  \begin{center}
    \scalebox{0.4}{\input{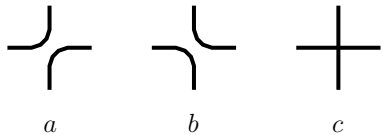}}
  \end{center}
  \caption{The three allowed vertices in the Brauer loop model, with weights $1,
1, x$.}
  \label{fig:vertices}
\end{figure}

We consider, on the square lattice, the {\em Brauer loop model}
defined by the three
vertices $a$, $b$ and $c$, shown in figure~\ref{fig:vertices}. The local
Boltzmann
weights are given by~:
\begin{equation}
\omega_a, \omega_b, \omega_c = 1, 1, x
\end{equation}
We give an additional weight $n$ for each closed loop.

\begin{figure}
  \begin{center}
    \scalebox{0.4}{\input{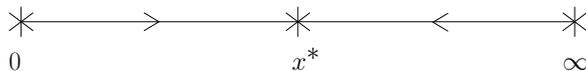}}
  \end{center}
\caption{Renormalization flow of the loop model, for the parameter $x$.}
\label{fig:flow-brauer}
\end{figure}

The renormalization flow for the parameter $x$ is shown in
figure~\ref{fig:flow-brauer}.
The fixed point $x=0$ corresponds, for $-2 \leq n \leq 2$, to the
low-temperature
phase of the $O(n)$ model \cite{nienhuis82}.
The fixed point $x^*$ attracts, for integer \cite{jrs03} and presumably
also real values of $n$, all the points in the
phase $x>0$. This phase contains an integrable point $x_{\mathrm{int}}=
(1-n/2)/2$
\cite{nienhuis98}. At this particular value of $x$, the $R$-matrix
(expressed in the
connectivity basis) satisfies the Yang-Baxter equations for any real value
of $n$.

The goal of this paper is to study the non-intersection properties of the
Brauer loop model. Recall that in the Brownian case, the simplest
such property is the disconnection exponent, which governs the probability 
that the origin of 
a single Brownian path remains accessible from infinity without the 
path being crossed. Let us recall some of the well established 
results in this case. If one considers 
a random walk of length $S$, the origin stays 
connected to infinity with probability $P(S)\sim S^{-\zeta_{1}}$, with
$\zeta_{1}=1/8$. One can also imagine growing the walk until it 
reaches the circle of radius $R$, and ask the probability with which 
it has left the origin connected to infinity, in which case $P(R)\sim 
R^{-2\zeta_{1}}$. Finally, one can consider a two-point correlation 
function defined as the  weighted sum \cite{DK88} over all walks going 
from the origin to a point $\mathbf{r}$ for which either extremity
remains connected to infinity. When the monomer fugacity equals the 
critical value, $G(\mathbf{r})\sim |\mathbf{r}|^{-2x_{1}}$ with 
$x_{1}=2\zeta_{1}=1/4$.

By analogy, we define an {\em interface} as
a line (of arbitrary length) that the Brauer loops are not allowed
to cross. In particular, we adress the following problems~:
\begin{enumerate}
\item Define the above loop model on a finite lattice of $N$ sites.
Let the origin $0$ be a point inside the lattice, and
$\mathbf{r}$ a point on the boundary of the lattice.
Let $P(\mathbf{r})$ be the probability that there exists an
interface going from the origin to the point $\mathbf{r}$. 
How does $P(\mathbf{r})$ decay when $r=|\mathbf{r}|$ goes to infinity ?
\item When this interface exists, it is, by definition, a self-avoiding
walk---for easy reference we henceforth refer to it as
$\Gamma$. The original loop system around $\Gamma$ acts like an environment,
affecting its geometrical properties. What is the universality class of
$\Gamma$ ?
\end{enumerate}

\section{Definition of a fluctuating interface}
\label{sec:def-gamma}

It turns out that  in the Brauer loop model with a single loop,
for any $x>0$, the probability $P(\vectr)$ behaves as~:
\begin{equation} \label{eq:p-exp}
P(\vectr) \sim \exp(-r/\xi)
\end{equation}
where $\xi$ is a length-scale depending on $x$. 
This behaviour is profundly different from the one of Brownian motion, 
where the probability 
$P(\vectr)$ for a long Brownian trajectory decays algebraically (a 
consequence of subadditivity arguments \cite{Lawlerbook}). This
indicates that, without modification, the interfaces in the Brauer 
model at $n=0$ do not have interesting scaling properties.

Let us see more explicitely why such behaviour arises.
Define three partition functions 
$Z_0(N)$, $Z'(N| 0, \vectr)$ and $Z_0(N|\Gamma)$ as follows.
\begin{itemize}
\item Let $Z_0(N)$ be the partition sum of the loop model for $N$ sites.
\item Let $Z'(N| 0, \vectr)$ be the partition sum of the loop model with the
constraint that there exists an interface not crossed by the loop, going
from $0$ to $\vectr$.
When a loop configuration admits an interface $\Gamma$, then this interface
is unique (or else there would be several loops). Thus, one can rewrite
the sum
$Z'(N| 0, \vectr)$ by grouping the terms corresponding to the same interface
$\Gamma$~:
\begin{equation}
Z'(N| 0, \vectr) = \sum_{\Gamma \in \SAW(0, \vectr)} Z_0(N|\Gamma)
\end{equation}
where $\SAW(0, \vectr)$ is the set of all self-avoiding walks
with ends fixed to $0$ and $\vectr$, and $Z_0(N|\Gamma)$ is
the partition sum of the loop model with the interface defined
by the self-avoiding walk configuration $\Gamma$.
\end{itemize}
Consider the presence of an effective interface energy in $Z_0(N|\Gamma)$~:
\begin{equation} \label{eq:Fs}
Z_0(N|\Gamma) \sim Z_0(N) \ \exp[ -l \times F_s(x) ]
\end{equation}
where $l$ is the length of $\Gamma$, and $F_s$ is a free energy per unit of
length of $\Gamma$.
The number of self-avoiding walk configurations of length $l$ is approximately
$(\mu_\SAW)^l$, where $\mu_\SAW=2.6381(5)$ \cite{jensen04} is the
connectivity constant of self-avoiding walks on the square lattice.
If the quantity $(F_s - \log \mu_\SAW)$ is positive, then the behaviour of
$Z'(N| 0, \vectr)$ is governed by configurations where the interface has
minimal length~:
\begin{eqnarray}
l & \simeq & r \\
Z'(N| 0, \vectr) & \sim & Z_0(N) \ \exp[ -(F_s-\log \mu_\SAW ) r]
\end{eqnarray}
This accounts for the exponential behaviour of $P(\vectr)=Z'(N| 0,
\vectr)/Z_0(N)$.

\begin{figure}
\begin{center}
\includegraphics[scale=0.2]{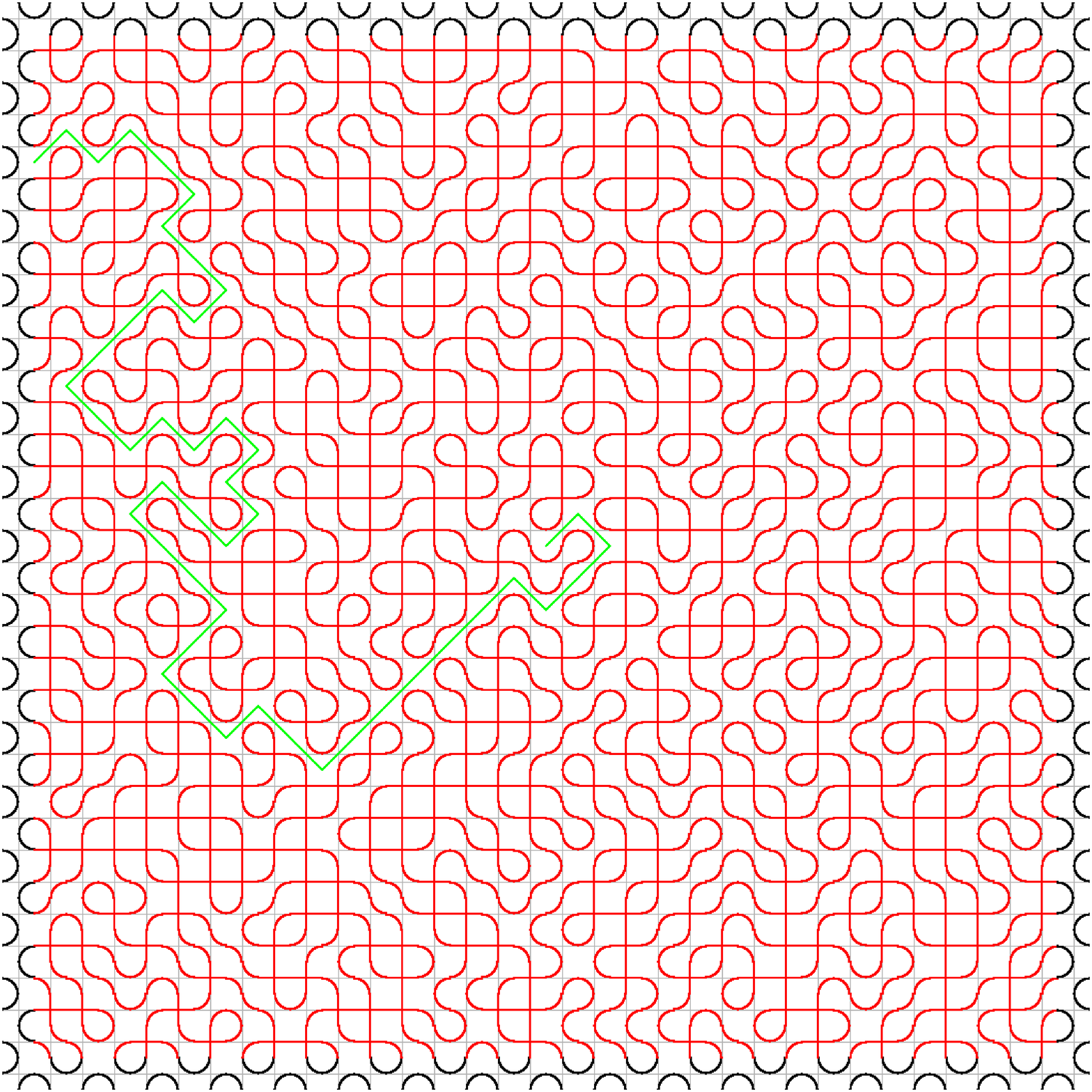}
\end{center}
\caption{A typical configuration contributing to the partition sum $Z(N|0,
\vectr)$.}
\label{fig:conf-L32-W-1.6}
\end{figure}

In order to define a scale-invariant interface, we  need to 
allow its length  $l$ to develop big fluctuations. A simple way to do
this is to introduce an additional
Boltzmann weight $w^l$, where $l$ is the length of $\Gamma$.
Thus, we are interested in the study of the partition function~:
\begin{equation} \label{eq:Z}
Z(N| 0, \vectr) \equiv \sum_{\Gamma \in \SAW(0, \vectr)} w^l Z_0(N|\Gamma)
\end{equation}
This function contains two external parameters~: $x$, the fugacity of a
crossing in
the loop model defined by $Z_0(N|\Gamma)$; and $w$, that controls the
length $l$ of the interface $\Gamma$.
A typical configuration contributing to the partition sum~\eqref{eq:Z}
is shown in figure~\ref{fig:conf-L32-W-1.6}.

\section{Phase diagram}
\label{sec:phase-diagram}

Using renormalization group arguments and
anticipating on our numerical and analytical results, we try to give
a consistent picture of the phase diagram of the model defined by
equation~\eqref{eq:Z}.

Let us fix the value of $x \geq 0$, and observe the system as we vary $w$.
In this picture, the interface $\Gamma$ is a self-avoiding walk, with
its self-interactions induced by the surrounding loop model.
According to equation~\eqref{eq:Fs}, we expect a phase transition at
$w^*(x)=\exp[F_s(x)] / \mu_\SAW$.

Note that it may be possible to check the previous relation. We
can compute numerically the value of $w^*$. If we could also compute
independently the value of $F_s$, then we would be able to compare these
values with the connectivity constant $\mu_\SAW$, which is a
well-known constant \cite{jensen04}.

\begin{figure}
  \begin{center}
    \scalebox{0.4}{\input{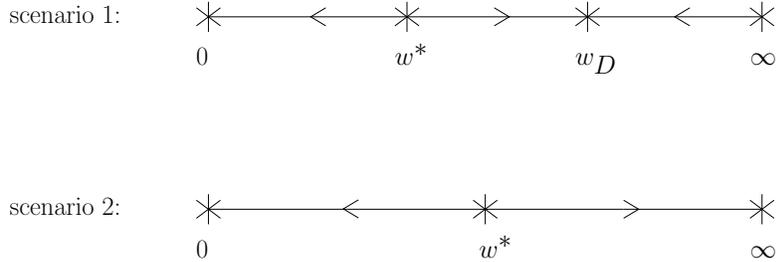}}
  \end{center}
  \caption{Two scenarii for the flow diagram of the parameter $w$.
Numerical studies
    suggest that scenario 1 is correct.}
  \label{fig:flow-w}
\end{figure}

Given $w^*$, the qualitative behaviour of the system as a function
of $w$ is expected to be as follows~:
\begin{itemize}

\item If $w < w^*$, the behaviour of $Z(N| 0, \vectr)$
is governed by configurations where the interface has minimal length~:
this phase is ``non-critical'' from the point of view of the interface.
The statement~\eqref{eq:p-exp} amounts to the
inequality~: $w^* > 1$, for every $x>0$.
In the phase $w < w^*$, the interface acts as a non-fluctuating boundary
for the loop model. So we expect the points of this phase to flow towards
one of the fixed points described in section~\ref{sec:loop-model}.

\item If $w = w^*$, the interface length $l$ fluctuates above its minimal
value $r$, with fluctuations of the order of the system size. It is a fractal
object, with Hausdorff dimension $1 < d_f \leq 2$. We call this point the
``$\Gamma$-dilute'' critical point. 

\item If $w > w^*$, two scenarii, illustrated in figure~\ref{fig:flow-w},
  are possible \textit{a priori}. In scenario 1, there exists a fixed
  point $w_D$ such that $w^* < w_D <\infty$, 
  and the whole phase flows towards it. In scenario 2, no such 
  fixed point exists, and the whole phase flows towards the point
  $w=\infty$. Numerical studies suggest that scenario 1 is correct.

\end{itemize}

 This behaviour with respect to $w$ is very similar to the one of 
ordinary self-avoiding walks when driven into the dense phase by a 
large monomer fugacity \cite{duplantier-saleur87}.
The interface $\Gamma$ plays the role of the single chain,
and the parameter $w$ is the monomer fugacity for the
chain $\Gamma$.

\begin{figure}
  \begin{center}
    \scalebox{0.4}{\input{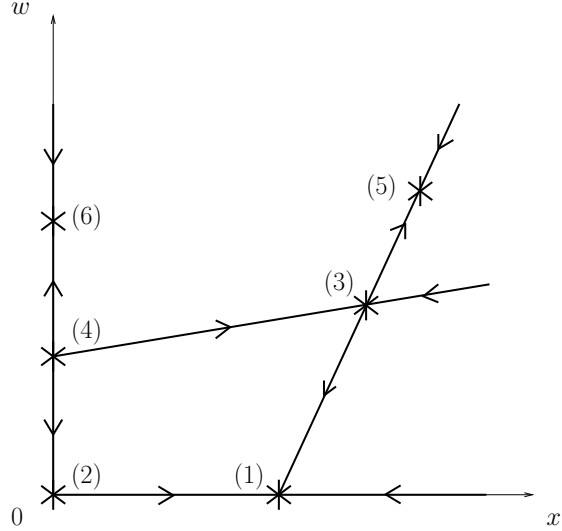}}
  \end{center}
\caption{Phase diagram of the model with $n=0$, in the $x-w$ plane.}
\label{fig:phase-diagram}
\end{figure}

To summarize, the phase diagram is given in figure~\ref{fig:phase-diagram}.
The points (1) and (2) are the critical points of the intersecting loop
model with
a fixed boundary. The point (3) (\textit{resp.} (4)) is the
``$\Gamma$-dilute'' point
for $x>0$ (\textit{resp.} $x=0$). The point (5) (\textit{resp.} (6))
is the ``$\Gamma$-dense'' point for $x>0$ (\textit{resp.} $x=0$).
The line passing through (3) and (4) has equation~: $w=w^*(x)$.

Let us focus on the critical point (4).
Since $x=0$, only vertices of type $a$ and $b$ are allowed,
and the point (4) is equivalent to the $Q$-state Potts model on the
square lattice, in a particular limit : 
in the Fortuin-Kasteleyn formulation, a
cluster has fugacity $Q \to 0$ and a bond has fugacity 
$v=\sqrt{Q}$. The leading term of the partition sum describes
equiprobable spanning trees. In this context, the 
interface~$\Gamma$ is exactly the set of \textit{red bonds} of the
spanning tree (the red bonds are the bonds of the spanning tree
which disconnect $0$ from $\mathbf{r}$ if removed). The fractal
dimension of this object is $5/4$ \cite{coniglio89, ds-prl87}.

The point (2) is just equivalent to the $Q$-state Potts model
in the limit described above, with a fixed cut line uncrossed
by the loop.

\section{Transfer matrix formulation}
\label{sec:transfer-matrix}

\subsection{Mapping from the plane onto the cylinder}
In order to estimate partition sums
such as $Z_0(N)$, $Z(N|0,\vectr)$ on a cylinder of fixed width $L$
and length $M \to \infty$, we compute the leading eigenvalues of the
transfer matrix. Denote them $\Lambda_L^{(0)}, \Lambda_L^{(1)},
\Lambda_L^{(2)}, \dots$
in decreasing order of modulus. Then the eigenvalue $\Lambda_L^{(i)}$
defines the
free energy density by surface unit $f_L^{(i)}$ in the following way~:
\begin{eqnarray}
Z_{\mathrm{cyl}}^{(i)} (M,L)& \sim & \left( \Lambda_L^{(i)} \right)^M \\
f_L^{(i)} = \frac{1}{ML} \log Z_{\mathrm{cyl}}^{(i)} (M,L)
  & \sim & \frac{1}{L} \log \Lambda_L^{(i)}
\end{eqnarray}
For conformally invariant models, finite-size corrections to $f_L^{(i)}$
give access
to the central charge $c$ and to the operator dimensions $X^{(i)}$
\cite{cardy86,cardy84}~:
\begin{eqnarray}
f_L^{(0)} & = & f_\infty + \frac{\pi c}{6 L^2} + o(L^{-2}) \\
f_L^{(0)} - f_L^{(i)} & = & \frac{2 \pi X^{(i)}}{L^2} + o(L^{-2})
\label{eq:XL}
\end{eqnarray}
where the dimensions are defined in plane geometry by~:
\begin{equation}
Z_{\mathrm{pl}}^{(i)} (0, \vectr) / Z_{\mathrm{pl}}^{(0)} \sim 1/r^{2X^{(i)}}
\end{equation}

\subsection{Transfer matrix for the loop model}
\begin{figure}
  \begin{center}
    \scalebox{0.4}{\input{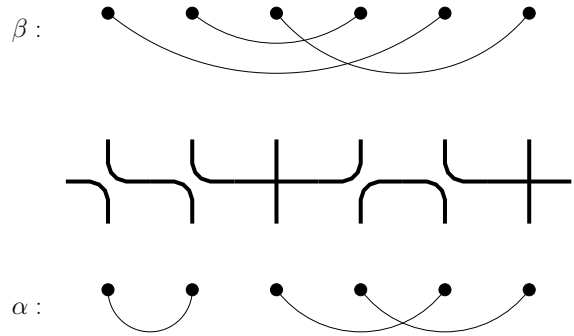}}
  \end{center}
  \caption{An example term in the matrix element $T_{\beta \alpha}$ of the
    transfer matrix.}
  \label{fig:transfer-matrix}
\end{figure}

Our convention is a transfer matrix $T_L$ acting in the vertical
direction, from bottom to top.
Let the width $L$ of the cylinder be an even integer.
Consider the strands living on the vertical edges of the lattice.
A \textit{row configuration} is a pairing of these strands. There are
$(L-1)!!=1 \times 3 \times 5 \times \dots (L-1)$
pairings of $L$ objects.
If $\alpha$ and $\beta$ are two row configurations, then the matrix
element $(T_L)_{\beta \alpha}$ is the total Boltzmann weight of $L$-vertex
configurations that map $\alpha$ to $\beta$ 
(see figure~\ref{fig:transfer-matrix}).
This Boltzmann weight includes a factor $n$ each time a loop is closed.

\begin{figure}
  \begin{center}
    \includegraphics[scale=0.3]{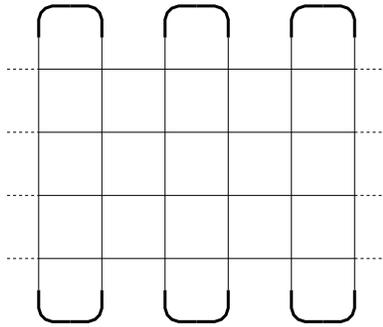}
  \end{center}
  \caption{Closing of the loop at the two boundaries of the
    strip. Dashed lines indicate periodic boundary conditions.}
  \label{fig:closing}
\end{figure}

On a cylinder of length $M$, the partition sum is given by~:
\begin{equation} \label{eq:Z_cyl}
Z_{0, \mathrm{cyl}}(M, L) =
\langle \alpha_{\mathrm{out}} | \left( T_L \right)^M | \alpha_{\mathrm{in}} \rangle
\sim \left( \Lambda_L^{(0)} \right)^M
\end{equation}
where $|\alpha_{\mathrm{in}} \rangle$ and $| \alpha_{\mathrm{out}} \rangle$ are two particular
vectors coding the closing of the loop at each extremity of
the cylinder. For example, one may choose to close the loop as
illustrated in figure~\ref{fig:closing}.

\subsection{Definition of an interface on the cylinder}

\begin{figure}
  \begin{center}
    \includegraphics[scale=0.5]{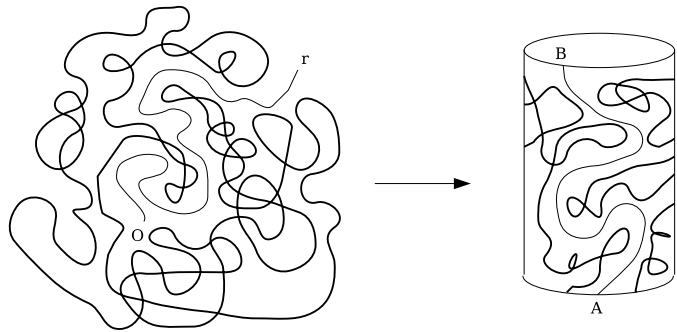}
  \end{center}
  \caption{Mapping of the interface problem from the plane to the cylinder.}
  \label{fig:cylindre}
\end{figure}

Let $\Gamma$ be an interface connecting a point $A$ on the
bottom circle of the cylinder to a point $B$ on the top circle of the
cylinder (see figure~\ref{fig:cylindre}).
Let $Z_{0, \mathrm{cyl}}(M, L|\Gamma)$ be the partition sum
of the loop model constrained by the interface $\Gamma$.

Define, as in equation~\eqref{eq:Z}, the partition 
sum~$Z_{\mathrm{cyl}}(M, L)$~:
\begin{equation}
Z_{\mathrm{cyl}}(M, L) \equiv \sum_{\Gamma} w^l Z_{0, \mathrm{cyl}}(M,
L|\Gamma)
\end{equation}

Note that the vector $\left( T_L \right)^M | \alpha_{\mathrm{in}} \rangle$
may contain components on configurations which admit several 
interfaces. This is because connectivity states are defined 
by unclosed loop segments. In contrast, the final scalar 
product~\eqref{eq:Z_cyl} contains only contributions from single
closed loop configurations admitting exactly one interface.

An analog observation can be done in the plane geometry. Suppose the
system is defined on a finite square of $N=L^2$ sites. A given loop
configuration contributing to the partition sum~$Z(N|0,\mathbf{r})$
admits one and only one interface $\Gamma$ going from $0$ to a point
of the boundary. But if one considers the subsystem contained in a
smaller square centered on $0$ and of side $L'<L$, one may find 
several interfaces going from the origin to a point of the smaller 
square.

\subsection{A local model for the interface weight}

\begin{figure}
  \begin{center}
    \scalebox{0.4}{\input{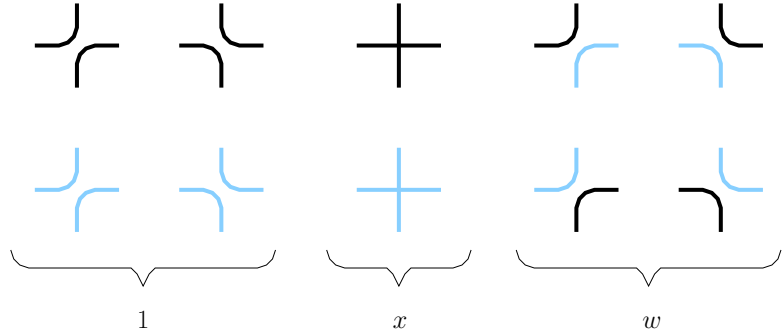}}
  \end{center}
  \caption{Vertex weights in the colored loop model.}
  \label{fig:vertex-col}
\end{figure}

The existence of the interface $\Gamma$ and its weight $w^l$ are
non-local features, which can fortunately be expressed by local
constraints on a colored loop model with loop fugacity $n=0$.
The loops may take two colors, according to the rules~:
\begin{itemize}
\item vertices of type (a) and (b) involve two strands of any colors,
  but vertices of type (c) must involve two strands of the same color.
\item the vertex weights are those given in figure~\ref{fig:vertex-col}.
\item a strand must change its color when it crosses a seam going along
  the cylinder.
\end{itemize}
The loop model defined by these rules satisfies the following properties~:
\begin{itemize}
\item Every colored loop configuration on the cylinder
  admits exactly one interface.
\item If a loop configuration admits a unique 
  interface, then it can be colored in exactly two ways.
\item The interface length $l$ is equal to the number of vertices
  which involve two different colors.
\end{itemize}
Thus the partition function of the colored loop model is equal (up to
a constant factor) to the partition function~$Z_{\mathrm{cyl}}(M,L)$.

\subsection{Other partition sums}

In the previous discussion we described the partition sum of one loop
with one interface. However, the colored loop model allows us to
compute partition sums with up to two interfaces, and/or containing
open trails.

The trick used to impose the presence of two interfaces is simply to
eliminate the color change across the infinite seam.

Open trails are introduced in a standard way into the connectivity
basis, by adding states where one or more points are connected to a
``virtual'' partner. The additional rule is that these points cannot
be connected with each other.

Denote $n_1$ the number of open trails and $n_2$ the number of closed
loops. The cylinder circumference $L$ must have the same parity as 
$n_1$. The partition sums which we are able to reach with the colored 
loop model have $n_1 + n_2 \le 2$ interfaces.
The corresponding physical exponent is denoted $X_{(n_1,n_2)}$.

\subsection{Numerical results}
\label{sec:num-transfer}

Our numerical procedure is to diagonalize the transfer matrix
of the colored loop model, using the power method.
Define the exponent $X_{(0,1)}$ by~:
\begin{equation}
  \frac{ Z_{\mathrm{pl}}(N|0, \vectr) }{ Z_{0,\mathrm{pl}}(N) }
  \sim \frac{ 1 }{ r^{2X_{(0,1)}} }
\end{equation}
Using equation~\eqref{eq:XL}, we define the finite-size estimates
$X_{(0,1)}(L)$ by~:
\begin{equation}
  X_{(0,1)}(L) \equiv \frac{L^2}{2\pi} \left( f_L^{(0)} - f_{L,(0,1)} \right)
\end{equation}

\begin{figure}
  \begin{center}
    \input{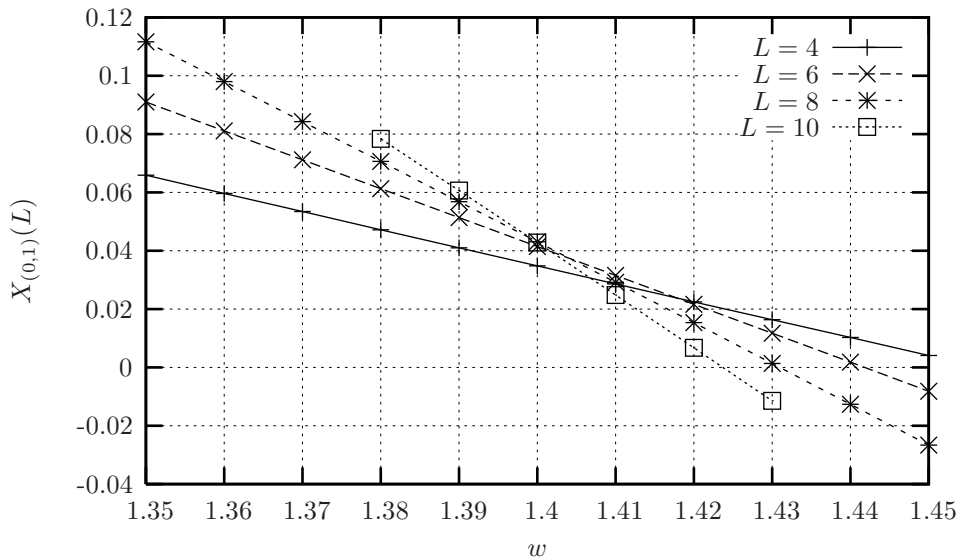}
  \end{center}
  \caption{Estimates of the exponent $X_{(0,1)}$ for $x=1$.}
  \label{fig:X1g}
\end{figure}

\begin{figure}
  \begin{center}
    \input{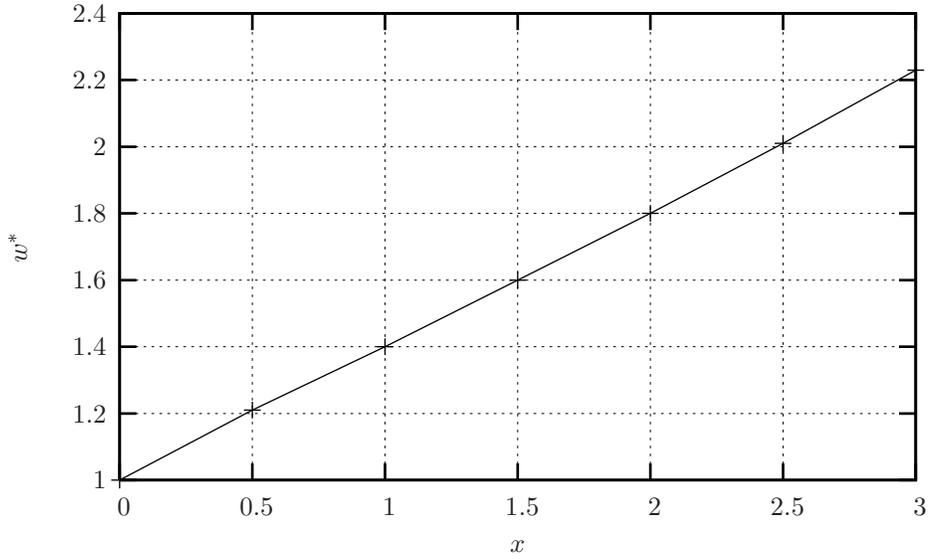}
  \end{center}
  \caption{Position of the critical point $w^*$ as a function of $x$. The
    values were obtained for width $L=8, 10$. The curve is very well 
    approximated by the formula~: $w^*=1+0.4 \ x$.}
  \label{fig:w-etoile}
\end{figure}

These estimates are also useful to determine the position of the critical
point $w^*$. Indeed, consecutive estimates $X_{(0,1)}(L)$ and
$X_{(0,1)}(L+2)$ coincide at $w=w^*(L, L+2)$, and---by analogy with
standard phenomenological renormalisation group methods---we expect 
this value to converge to $w^*$
when $L$ is large. This method is illustrated in figure~\ref{fig:X1g}.
The resulting critical line is plotted in figure~\ref{fig:w-etoile}.

Following a similar method, we define the general exponent
$X_{(n_1, n_2)}$ by comparing the free energy $f_{L,(n_1,n_2)}$ with
the free energy associated with the greatest eigenvalue for the same
system width~:
\begin{equation}
  X_{(n_1,n_2)}(L) \equiv \frac{L^2}{2\pi} \left( f_L^{(0)} - f_{L,(n_1,n_2)} \right)
\end{equation}
Note that $f_L^{(0)}$ is the free energy of one
unconstrained loop ({\it resp.} trail) when $L$ is even ({\it resp.} odd).

The errorbars are given by the spacing of the consecutive
intersections of the finite-size
estimates. The results are compared with the empirical formula~:
\begin{equation} \label{eq:X-m}
  X_{(n_1,n_2)}=\frac{m^2-1}{12}, \qquad
  m=\frac{3}{2}n_1 + 2n_2 - 1
\end{equation}
to be discussed in the last section. 

\begin{table}
  \begin{center}
    \begin{tabular}{llll}
      Name & Description & Numerical & Formula~\eqref{eq:X-m}  \\
      \hline \\
      $X_{(0,1)}$ & 1 loop, 1 interface & $0.04 \pm 0.005$  & 0  \\
      $X_{(1,0)}$ & 1 trail, 1 interface & $0.00 \pm 0.005$ & -0.06 \\
      $X_{(0,2)}$ & 2 loops, 2 interfaces & $0.80 \pm 0.02$  & 0.67  \\
      $X_{(1,1)}$ & 1 loop, 1 trail, 2 interfaces & $0.50 \pm 0.008$ & 0.44\\
      $X_{(2,0)}$ & 2 trails, 2 interfaces & $0.21 \pm 0.02$ & 0.25
    \end{tabular}
    \caption{Estimates of the exponents $X_{(n_1,n_2)}$ by
      transfer-matrix diagonalisation.}
    \label{table:X}
  \end{center}
\end{table}

One must consider these results cautiously, because in the loop model,
logarithmic corrections appear in quantities such as $f_L^{(0)}$.

\section{Monte-Carlo simulation on a square}
\label{sec:MC}

In order to access directly some geometrical features of the model,
we simulate it by a Monte-Carlo procedure \cite{binder88} on an 
$L \times L$ square lattice.
First, we describe a Metropolis procedure used for the loop model defined
by the partition sum $Z_0(N)$. Then, we modify slightly this procedure to
simulate the ensemble defined by $Z(N|0, \vectr)$.
We discuss the main aspects of the algorithm~: detailed balance, ergodicity
and autocorrelation times. At the end, we give some numerical results
obtained by this method.

\subsection{Metropolis procedure for the loop model}

The statistical ensemble of interest is the set of loop configurations
consisting of \emph{exactly one loop} on an $L \times L$ square lattice.
Each loop configuration $\mathcal{L}$ is given the Boltzmann weight
$\Pi(\mathcal{L})=x^{N_c}$, where $N_c$ is
the number of crossings of the loop with itself.
The boundary of the square consists of ``frozen'' vertices, as shown in
figure~\ref{fig:conf-L32-W-1.6}.

The idea of the algorithm is to modify locally the system without violating
the one-loop constraint. Let us denote $P_{ij}$ the $2 \times 2$
``plaquette''
consisting of the vertices $\{ (i,j), (i+1,j), (i,j+1), (i+1,j+1) \}$, and
$\beta_{ij}$ the state of these four vertices. Every plaquette has eight
``external legs''. Outside $P_{ij}$, the loop connects these legs with each
other. Fixing this external connectivity---call it $\alpha$---defines
the set $B_\alpha$ of plaquette configurations $\beta$ respecting the
one-loop
constraint when adjoined with $\alpha$.

An elementary iteration consists of four steps~:
\begin{enumerate}
\item Pick uniformly a plaquette $P$ among $(L-1) \times (L-1)$
  possibilities.
  The state of the plaquette is denoted $\beta$.
\item Visit the loop to determine the external connectivity $\alpha$ of
  the legs of $P$.
\item Pick uniformly a configuration $\beta'$ from $B_{\alpha}$.
\item Perform the local change $\beta \to \beta'$ on $P$ with probability~:
  \begin{equation}
    W(\beta \to \beta') =
    \begin{cases}
      \frac{\Pi(\beta')}{\Pi(\beta)} & \text{if} \ \Pi(\beta') < \Pi(\beta) \\
      1 & \text{otherwise}
    \end{cases}
  \end{equation}
\end{enumerate}

This algorithm satisfies the detailed balance condition for the probability
distribution $\Pi(\mathcal{L})$ on the set of one-loop configurations~:
\begin{equation}
  \forall \mathcal{L}, \mathcal{L'} \quad
  \Pi(\mathcal{L}) W(\mathcal{L} \to \mathcal{L'})
  = \Pi(\mathcal{L'}) W(\mathcal{L'} \to \mathcal{L})
\end{equation}

\subsection{Metropolis procedure for the loop model with an interface}

Consider next the statistical ensemble of one-loop configurations
admitting an
interface $\Gamma$ that connects the center $O$ of the square to a point
$\vectr$ on the boundary of the square (see figure~\ref{fig:conf-L32-W-1.6}).
Each loop configuration is given
the weight $\Pi(\mathcal{L}) = x^{N_c} w^l$, where $N_c$ is the number of
crossings of the loop
with itself, and $l$ is the length of the interface $\Gamma$.
The above Metropolis procedure can be adapted to simulate this ensemble,
and is defined
by the elementary iteration~:
\begin{enumerate}
\item Pick uniformly a plaquette $P$ among $(L-1) \times (L-1)$
  possibilities.
  The state of the plaquette is denoted $\beta$.
\item Visit the loop to determine the external connectivity $\alpha$ of
  the legs of $P$.
\item Pick uniformly a configuration $\beta'$ from $B_{\alpha}$. This
  defines a new loop configuration.
\item Test whether the new loop configuration still admits an interface.
  If it is the case, denote $l'$ the length of the new interface.
  If not, reject the step and skip to a new iteration.
\item Perform the local change $\beta \to \beta'$ on $P$ with probability~:
  \begin{equation}
    W(\beta \to \beta') =
    \begin{cases}
      \frac{x^{N_c(\beta')} w^{l'}}{x^{N_c(\beta)}  w^{l}}
      & \text{if} \ x^{N_c(\beta')}  w^{l'} < x^{N_c(\beta)}  w^{l} \\
      1 & \text{otherwise}
    \end{cases}
  \end{equation}
\end{enumerate}

\begin{figure}
  \begin{center}
    \includegraphics[scale=0.4]{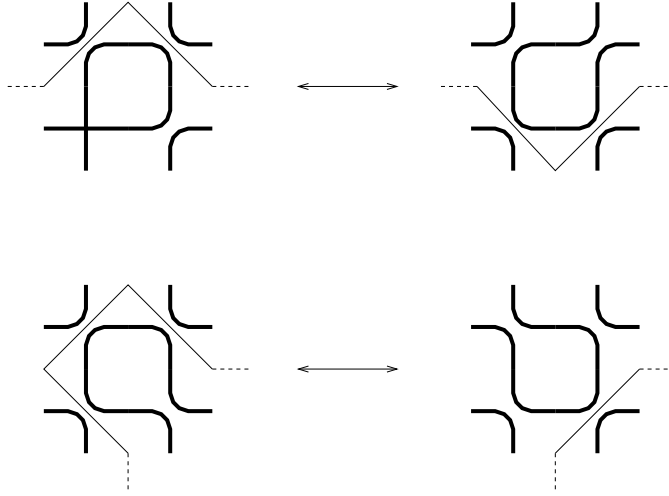}
  \end{center}
  \caption{Two examples of plaquette flips which alter locally the
    interface $\Gamma$ ($\Gamma$ is represented as a thin line).}
  \label{fig:metropolis}
\end{figure}

This algorithm is able to modify locally the interface, when the plaquette
picked at step 1 is close to the interface. Examples of this process
are given in figure~\ref{fig:metropolis}.

\subsection{Characteristic times}

\begin{figure}
  \begin{center}
    \input{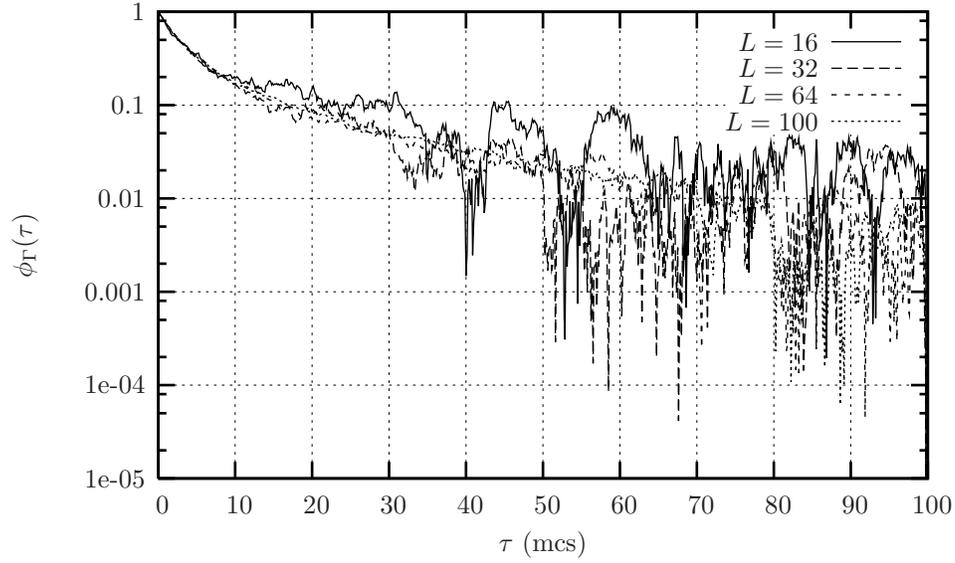}
  \end{center}
  \caption{Auto-correlation function $\phi_{\Gamma}(\tau)$
    for parameters $x=1$, $w=2$. The initial time was set to 
    $k_0=100 \ \mathrm{mcs}$.}
  \label{fig:phi_gamma}
\end{figure}

The Metropolis algorithm described above defines a Markov chain on the
set of one-loop configurations with an interface :~
$(\mathcal{L}_0, \mathcal{L}_1, \dots \mathcal{L}_K)$.
We are interested in the calculation of an observable~:
\begin{equation}
  \langle A \rangle =
  \frac{\sum_{\mathcal{L}} \Pi(\mathcal{L}) A(\mathcal{L})}
  {\sum_{\mathcal{L}} \Pi(\mathcal{L})}
\end{equation}
where the sums are over one-loop configurations with an interface.
The Markov chain is used to build an estimator for this quantity~:
\begin{equation} \label{eq:Abar}
  \bar{A}_{k_0, Q, \tau} =
  \frac{1}{Q} \sum_{q=0}^{Q-1} A \left( \mathcal{L}_{k_0+q \tau} \right)
\end{equation}
If the configurations $\mathcal{L}_k$ were independent and obeyed the
equilibrium distribution $\Pi(\mathcal{L})$, then the first two moments
of the estimator $\bar{A}_{k_0, Q, \tau}$ would be~:
\begin{eqnarray}
  \langle \bar{A}_{k_0, Q, \tau} \rangle & = & \langle A \rangle \\
  \langle \left[ \bar{A}_{k_0, Q, \tau} - \langle A \rangle\right]^2 \rangle
  & = & \frac{1}{Q} \langle \left[ A - \langle A \rangle\right]^2 \rangle
\end{eqnarray}
where the brackets denote averaging with respect to the distribution
$\Pi(\mathcal{L})$.
The parameters introduced in equation~\eqref{eq:Abar} play the following
roles~:
\begin{itemize}
\item The initial time $k_0$ is the number of configurations to discard at
  the beginning of the Markov chain. Indeed, the first configurations do not
  obey the equilibrium distribution $\Pi(\mathcal{L})$. The value of $k_0$ must be
  greater than the \textit{equilibration time} of the observable $A$ under the process.
\item The time interval $\tau$ allows us to discard redundant information
  arising from non-independent configurations $\mathcal{L}_k$, $\mathcal{L}_{k+1}$,
  \textit{etc} \cite{sokal96}.
  One has to choose a value $\tau$ greater than the
  \textit{auto-correlation time} of the observable $A$ under the process.
\item The number of samples $Q$ controls the accuracy of the estimator.
  If the interval $\tau$ is large enough, the error
  on $\langle A \rangle$ is proportional to $Q^{-1/2}$.
\end{itemize}
In order to determine the equilibration time and the auto-correlation time
for the observables of interest, we make a specific hypothesis. The observables
we want to compute depend only on the interface.
We define the ``interface overlap function''~:
\begin{equation}
  S_{\Gamma} (\mathcal{L}, \mathcal{L}') =
  2 \times \frac{\mathrm{length \ of \ } (\Gamma \cap \Gamma')} {l+l'}
\end{equation}
The hypothesis is that characteristic times for observables depending only on
the interface are bounded by a typical timescale associated with the decay
of $S_{\Gamma} (\mathcal{L}_0, \mathcal{L}_\tau)$ as a function of $\tau$.
This decay is well described by the auto-correlation function~:
\begin{equation}
  \phi_{\Gamma}(\tau) = \frac
  {S_{\Gamma} (\mathcal{L}_0, \mathcal{L}_\tau)
    - S_{\Gamma} (\mathcal{L}_0, \mathcal{L}_K)}
  {1 - S_{\Gamma} (\mathcal{L}_0, \mathcal{L}_K)} \ , \quad K \to \infty
\end{equation}
This function is plotted in figure~\ref{fig:phi_gamma}.
The time unit used for plotting is one Monte-Carlo sweep (mcs), corresponding
to $L^2$ elementary iterations of the Metropolis algorithm.

\subsection{Ergodicity}

The Markov chain $(\mathcal{L}_0, \mathcal{L}_1, \dots)$
defined by the Metropolis procedure converges to the equilibrium
distribution $\Pi(\mathcal{L})$ under two conditions~: detailed
balance and ergodicity. A Markov process is said ergodic if
and only if for any pair of configurations $(\mathcal{L}, \mathcal{L'})$,
there exists a path of transitions, going from $\mathcal{L}$ to
$\mathcal{L}'$, with non-zero transition probability~:
\begin{equation}
  \mathcal{L} \to \mathcal{L}_1 \to \mathcal{L}_2
  \to \dots \mathcal{L}_k \to \mathcal{L}'
\end{equation}
For the two versions of the algorithm, we do not have any proof of
ergodicity for a system of arbitrary size. Nevertheless, we have been
able to check exactly the ergodicity property for systems
of size $L=2, 4$. The transfer-matrix algorithm is
used to count the number of allowed configurations for small width $L$.
The Metropolis algorithm is modified to visit every configuration
obtained by a sequence of local changes, starting from an arbitrary 
configuration.
We find that every allowed configuration is reached by
the process. Table~\ref{table:ergodicity} gives the number of 
one-loop configurations on a $L \times L$ square, for the two cases.
 
\begin{table}
  \begin{center}
    \begin{tabular}{lllc}
      $L$ & $\mathcal{N}_{1 \, \mathrm{loop}}$ &
      $\mathcal{N}_{1 \, \mathrm{loop}, 1 \, \mathrm{interface}}$ & ergodicity
      checked \\
      \hline \\
      2 & 40 & 32 & $\times$ \\
      4 & 5 373 952 & 3 072 000 & $\times$ \\
      6 & 4 380 037 227 741 184 & 1 705 236 234 240 000 & \
    \end{tabular}
  \end{center}
  \caption{Ergodicity check of the Metropolis algorithms 
    for small system sizes.}
  \label{table:ergodicity}
\end{table}

\subsection{Numerical results}

\begin{figure}
  \begin{center}
    \input{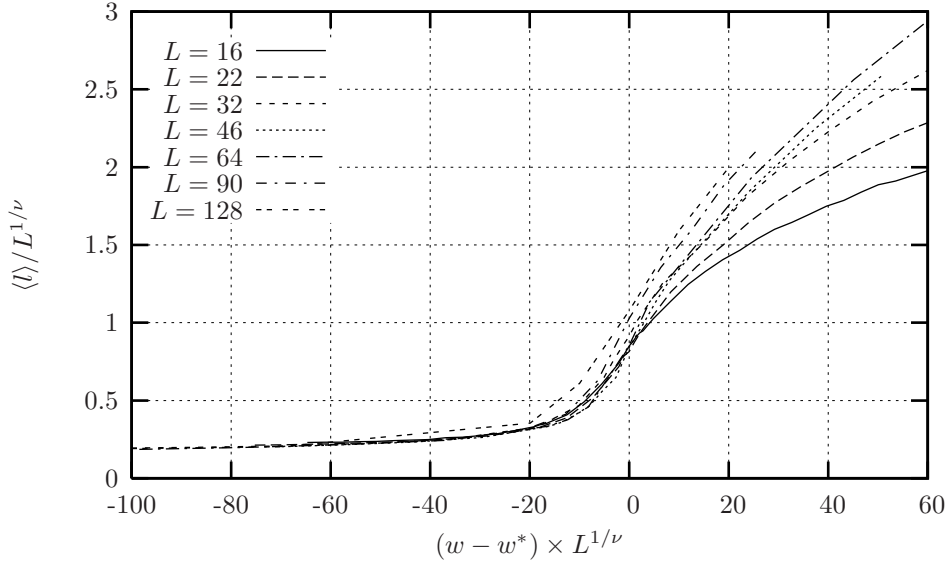}
  \end{center}
  \caption{Mean length $\langle l \rangle$ of the interface, as a function
    of $w$, for $x=1$.
    The parameters $\{w^*(L)\}$ and $\nu$ are determined by the maximum of
    $\langle \Delta l^2 \rangle$ (see figure~\ref{fig:fluct-max}).}
  \label{fig:scaling-l}
\end{figure}

\begin{figure}
  \begin{center}
    \input{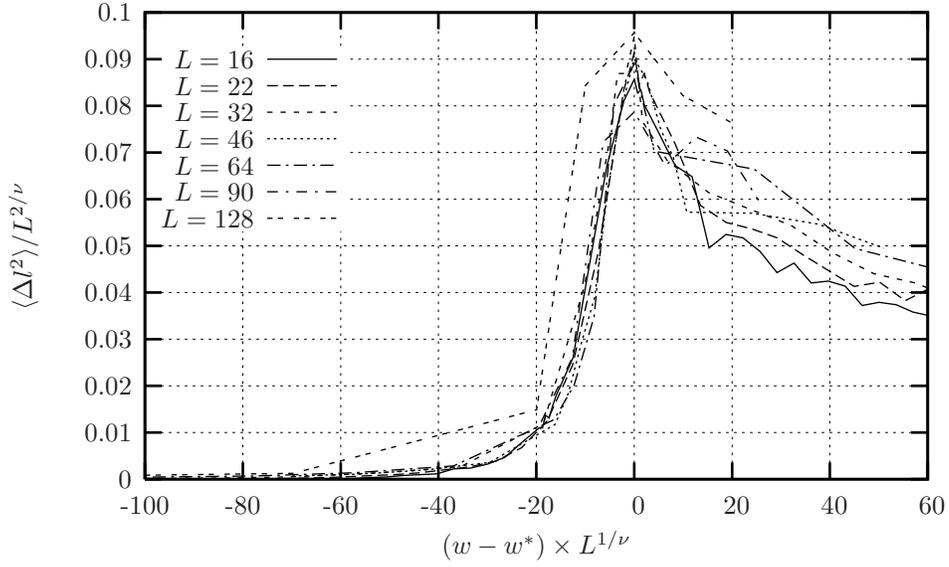}
  \end{center}
  \caption{Fluctuations of the length $l$ of the interface, as a function
    of $w$, for $x=1$.
    The parameters $\{w^*(L)\}$ and $\nu$ are determined by the maximum of
    $\langle \Delta l^2 \rangle$ (see figure~\ref{fig:fluct-max}).}
  \label{fig:scaling-Dl2}
\end{figure}

\begin{figure}
  \begin{center}
    \input{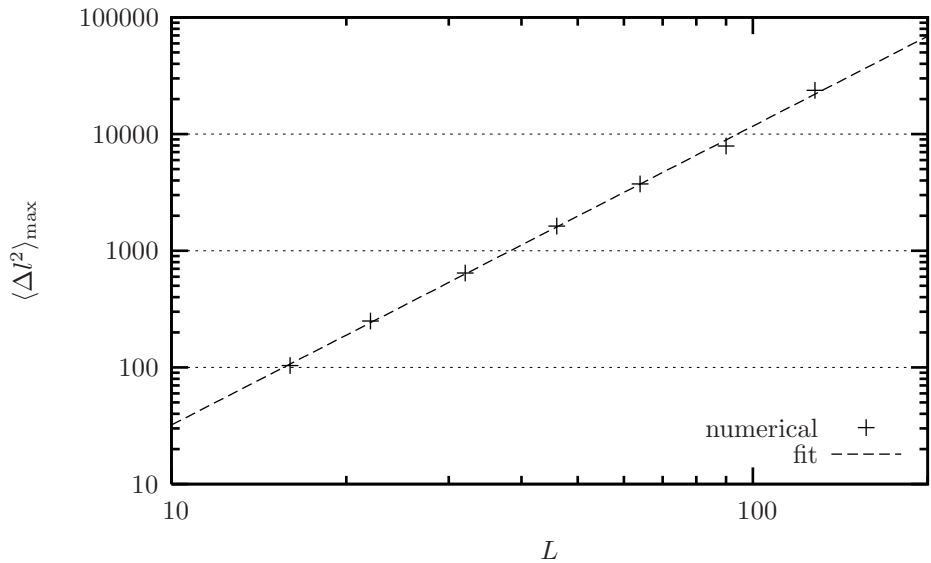}
  \end{center}
  \caption{Maximum value of the fluctuations of the length $l$ of the
    interface,
    as a function of the system size $L$. The linear fit gives
    $\langle \Delta l^2 \rangle_{\mathrm{max}} \propto L^{2/{\nu}}$, with
    $1/{\nu}=1.28$.}
  \label{fig:fluct-max}
\end{figure}

Using Monte-Carlo simulations for various system sizes $L=16,\dots,128$,
with the parameters~:
\begin{eqnarray}
  k_0 &=& 100 \ \mathrm{mcs} \notag \\
  \tau &=& 100 \ \mathrm{mcs} \notag \\
  Q &=& \bigg\{ \begin{array}{ll} 
    1000 \ \mathrm{samples} & \mathrm{for} \ L<128 \\
    100  \ \mathrm{samples} & \mathrm{for} \ L=128 
  \end{array} \notag
\end{eqnarray}
we estimate the average length $\langle l \rangle$ of the interface
$\Gamma$, and its fluctuations. We obtain an evidence 
(see figures~\ref{fig:scaling-l} and \ref{fig:scaling-Dl2})
for the scaling laws~:
\begin{eqnarray}
  \langle l \rangle & = & L^{1/{\nu}} H_1 \left( [w-w^*(L)] L^{1/{\nu}}
  \right) \\
  \langle \Delta l^2 \rangle & = & \langle l^2 \rangle - \langle l \rangle^2
  =  L^{2/{\nu}} H_2 \left( [w-w^*(L)] L^{1/{\nu}} \right)
\end{eqnarray}
where $w^*(L)$ is the ``finite-size critical point'', and
$H_1, H_2$ are scaling functions. The function $H_1(u)$ is increasing on
the real line,
and the function $H_2(u)$ has a maximum at $u=0$.
We define $w^*(L)$ as the value of the parameter $w$ which
maximizes the fluctuation amplitude $\langle \Delta l^2 \rangle$.
The maximum of the quantity $\langle \Delta l^2 \rangle$ is expected
to follow the scaling law~:
\begin{equation}
  \langle \Delta l^2 \rangle_{\mathrm{max}} \propto L^{2 / \nu}
\end{equation}
This scaling law is confirmed by the numerical simulation,as shown in 
figure~\ref{fig:fluct-max}. The numerical fit leads to~:
\begin{equation} \label{eq:nu}
  1 / \nu = 1.28 \pm 0.018
\end{equation}

\subsection{Scaling laws for the interface}
\label{sec:scaling}

We interpret the scaling laws for the length $l$ of the interface,
using the scaling hypothesis for the partition function, close
to the critical point $w=w^*$.
Suppose that the partition function $Z(N|0, \vectr)$ is a
two-point correlation function in some critical field theory.
In this theory, the correlation length $\xi$ is related to the
deviation from the critical point by~:
\begin{equation}
  \xi \propto |w-w^*|^{-\nu}
\end{equation}
The function $Z(N|0, \vectr)$ defines the anomalous dimension $\eta$~:
\begin{equation}
  Z(N|0, \vectr) = \frac{1}{r^{d-2+\eta}} g(r/\xi)
\end{equation}
where $d=2$ is the space dimension, and $g$ is a scaling function.
The partition function $Z(N|0, \vectr)$ is also the generating function
for the length $l$ of the interface going from $0$ to $\vectr$~:
\begin{eqnarray}
  \langle l \rangle &=& \frac{\partial \log Z(N|0, \vectr)}{\partial \log w}
  = r^{1/{\nu}} h_1(r/\xi) \\
  \langle \Delta l^2 \rangle &=& \frac{\partial^2 \log Z(N|0,
    \vectr)}{\partial (\log w)^2}
  = r^{2/{\nu}} h_2(r/\xi)
\end{eqnarray}
where the scaling functions $h_1, h_2$ can be expressed in terms of the
function $g$.
So, in particular, the fractal dimension of the interface at the critical
point is $d_f=1/{\nu}=1.28$.

This result can be compared with the transfer-matrix calculations of
the exponents $X_{(n_1,n_2)}$. As for the case of polymers
\cite{nienhuis82, duplantier-saleur87}, the thermal exponent $\nu$
is related to the ``two-leg exponent'' $X_{(0,2)}$~:
\begin{equation}
  X_{(0,2)} = 2 - \frac{1}{\nu}
\end{equation}
Indeed, the exponent~$X_{(0,2)}$ corresponds to the insertion of 
two interfaces. According to this argument, the estimated value of
$X_{(0,2)}$ (see table~\ref{table:X}) gives a fractal dimension~:
\begin{equation}
  d_f = 1.2 \pm 0.02
\end{equation}
This estimate of $\nu$ is only marginally compatible with 
\eqref{eq:nu}, but this was to be expected
given the difficulty in estimating reliable error 
bars within the transfer matrix method.

\section{Conclusion}
\label{sec:conclusion}

In Brownian motion, the fractal dimension of the boundary is known to 
be $D_{f}=1/\nu={4/3}$, a celebrated conjecture of Mandelbrot's, 
established rigorously by Lawler, Schramm and Werner \cite{LSW01}.%
\footnote{We note that in this case there is no microscopic duality 
 between the non-intersecting  Brownian walk and any sort of  ``interface'', 
 since the Brownian walk is not forced to fill up space. It would be 
 interesting to see if any such interface can be defined.}
At the critical value of $w$ where the 
interface is allowed to grow in our fully packed trail model, the 
result~\eqref{eq:nu}
is quite close to this value. Since by construction $\Gamma$ is 
self avoiding, and since few universality classes are known for 
self-avoiding walks (even though the background of trails induces non 
local interactions in $\Gamma$) it is tempting to speculate that 
our $\nu=3/4$ indeed. 

In the context of Brownian motion, the problem closest to ours is the 
problem of non-intersections of packets of walks. We shall restrict 
to situations with packets made of one or two walks only. Suppose we 
have $n_{1}$ walks and $n_{2}$ pairs of walks, and suppose we demand that no 
intersection occurs between either sets, while for each pair of walks 
in the set $n_{2}$, intersections are allowed. The exponent is then 
known to be \cite{LSW01}~:
\begin{equation} \label{eq:dupl}
  x=2\zeta(n_{1},n_{2})=\frac{m^{2}-1}{12}
\end{equation}
where~: 
\begin{equation}
  m=2n_{1}+3n_{2}
\end{equation}
One recovers the usual case of non-intersection exponents when 
$n_{2}=0$, $n_{1}\equiv L$ in Duplantier's notations \cite{Duplantier98}
(not to be confused with $L$ the system size in this paper).

Assuming (without clear justification) that similar properties might hold 
for the non-intersection exponents of trails, we found that 
the following empirical formula~:
\begin{equation}
  x = \frac{m^2-1}{12}, \qquad
  m = \frac{3}{2} n_1 + 2 n_2 - 1
\end{equation}
gives a decent fit to our data, as mentioned in section~\ref{sec:num-transfer}. 

Recall that the numerical data used to obtain this conjecture are
affected by very bad convergence properties. An analytical
approach and further numerical confirmation are needed to make 
progress on the problem. For example, one may think of 
attacking the problem  using quantum gravity methods.

An important point concerns the value of $n$ (the loop fugacity) that
is best suited for comparison with Brownian motion. We chose in this
paper $n=0$ mostly by analogy with the usual SAW case, but notice that
formally the corresponding central charge is $c=-1$ \cite{jrs03}. This
might not be the best choice, since non-intersection exponents such as
\eqref{eq:dupl} appear in a CFT with $c=0$ \cite{DK88,Duplantier98}. On
the other hand, the value $c=-1$ is really a property of the ``bulk''
of dense trails, and it is not excluded that their frontiers might be
described by a different CFT. In fact, it might well be that
non-intersection exponents are independent on the value of $n$---after
all, the dense trail models with $n<2$ have properties which, in many
respects, are independent of $n$ (such as the fuseau exponents, which
are all vanishing \cite{jrs03}).  Some thoughts on the proper 
generalization of the lattice model needed to deal with the case 
$n \neq 0$ are presented in the Appendix below.
We hope to get back to this question soon.

\vspace{0.5cm}

\noindent
{\bf Acknowledgements}\\ One of the authors (YI) thanks B. Nienhuis
and W. Kager for their hospitality at the ITFA, and for interesting
discussions on generalizations to $n \neq 0$.  Discussions with 
M. Bauer, B. 
Duplantier  and S. Majumdar (from whom we learned about 
the known results on the point (4) of
figure~\ref{fig:phase-diagram}) are also gratefully acknowledged.

\section*{Appendix : Ideas for generalizations to $n \neq 0$}
\label{sec:generalizations}

In the planar geometry, when $n$ is non-zero, several loops are allowed
to cover the lattice, so the interface defined in
section~\ref{sec:def-gamma} is
no longer unique.

In the case $n=0$, the coloring of loops with two colors was used
in section~\ref{sec:transfer-matrix} to express with local weights
the existence of an interface. When $n$ is non-zero, this trick
can inspire some generalizations of the problem. These are best
defined on a cylinder.

Let $Z_{\mathrm{col}}$ be the partition sum of the model defined
by the vertices weights of figure~\ref{fig:vertex-col}, and the
non-local weight $n$ for each closed loop.
Given a particular configuration $\mathcal{L}$ consisting of several
uncolored loops, the model allows several colorings 
of $\mathcal{L}$. Intersections
among the loops of $\mathcal{L}$ define a partition into \textit{packets}
of mutually intersecting loops. Packets may be colored independently
in two ways, but loops belonging to the same packet must have
the same color. Thus, the number of such objects appears in the
partition sum~:
\begin{equation} \label{eq:Z_col}
  Z_{\mathrm{col}}(M, L) =
  \sum_{\mathcal{L}} \left( \prod_i w_i \right) n^{\# \mathrm{loops}} \ 
  2^{\# \mathrm{packets}}
\end{equation}
where the sum is over uncolored loop configurations, and the first factor
stands for the local weights of figure~\ref{fig:vertex-col}.

If a change of colors is imposed when the boundary is crossed, then no packet
is allowed to wrap around the cylinder. This is equivalent to the
existence of an
interface as defined in figure~\ref{fig:cylindre}. So, like in the case
$n=0$, this
model is able to express locally the existence of an interface.
Unfortunately,
we have not been able to define a unique interface, so we cannot control
its length $l$ like in the previous case.

A special case is when $x=0$. Then a packet is identical to a loop, and
the partition
sum becomes~:
\begin{equation}
  Z_{\mathrm{col}}(M, L) =
  \sum_{\mathcal{L}} \left( \prod_i w_i \right) (2n)^{\# \mathrm{loops}}
\quad (x=0)
\end{equation}
Note that this model contains the Potts model as a particular case,
when $w=1$. \\
\\

The form~\eqref{eq:Z_col} of the partition sum of the colored loop model
suggests
another generalization of the previous models, obtained by introducing a
second
non-local weight $m$ for each packet of closed loops~:
\begin{equation} \label{eq:Y}
  Y(x, n, m) =
  \sum_{\mathcal{L}} x^{\# \mathrm{crossings}} \  n^{\# \mathrm{loops}} \ 
  m^{\# \mathrm{packets}}
\end{equation}
where the sum is over uncolored loop configurations. \\
\\
\textit{NB} : In section~\ref{sec:def-gamma}, we saw that when $(n,m) =
(0,2)$, the model with
an interface is not critical. However, it may be possible to define a
limit $n \to 0$ for
the partition sum~\eqref{eq:Y}, which defines a critical model with an
interface.


\newpage
\begin{thebibliography}{99}

\bibitem{JS06} J.L. Jacobsen and H. Saleur,
  {\em Conformal boundary loop models}, math-ph/0611078.

\bibitem{LSW01} G.F. Lawler, O. Schramm and W. Werner,
  Acta Math. {\bf 187}, 237--273 (2001); math.PR/9911084.

\bibitem{Duplantier98} B. Duplantier,
  Phys. Rev. Lett. {\bf 81}, 5489 (1998).

\bibitem{Korchemsky} D.E. Derkachov, G.P. Korchemsky, A.N. Manashov, 
JHEP 0310 (2003) 053, and references therein.

\bibitem{DK88} B. Duplantier and K.-H. Kwon,
  Phys. Rev. Lett. {\bf 61}, 2514--2517 (1988).

\bibitem{Cao} M.S. Cao and E.G.D. Cohen,
 J. Stat. Phys. {\bf 87}, 147 (1997); cond-mat/9608159.

\bibitem{jrs03} J.L. Jacobsen, N. Read and H. Saleur,
  Phys. Rev. Lett. {\bf 90}, 090601 (2003); cond-mat/0205033.

\bibitem{nienhuis98} M.J. Martins, B. Nienhuis and R. Rietman,
  Phys. Rev. Lett. {\bf 81}, 504--507 (1998); cond-mat/9709051.

\bibitem{KN06} W. Kager and B. Nienhuis,
  J. Stat. Mech., P08004 (2006); cond-mat/0606370.

\bibitem{nienhuis82} B. Nienhuis,
  Phys. Rev. Lett. \textbf{49}, 1062 (1982).

\bibitem{Lawlerbook} G. F. Lawler, {\em Intersections of random 
  walks} (Birkh\"auser, Boston, 1996).

\bibitem{jensen04} I. Jensen,
  J. Phys. A \textbf{37}, 5503--5524 (2004).

\bibitem{duplantier-saleur87} B. Duplantier and H. Saleur,
  Nucl. Phys. B \textbf{290}, 291 (1987).

\bibitem{coniglio89} A. Coniglio,
  Phys. Rev. Lett. \textbf{62}, 3054 (1989).

\bibitem{ds-prl87} B. Duplantier and H. Saleur,
  Phys. Rev. Lett. \textbf{58}, 2325 (1987).

\bibitem{cardy86} H.W.J. Bl\"ote, J.L. Cardy and M.P. Nightingale,
  Phys. Rev. Lett. \textbf{56}, 742 (1986);
  I. Affleck, Phys. Rev. Lett. \textbf{56}, 746 (1986).

\bibitem{cardy84} J.L. Cardy,
  J. Phys. A {\bf 17}, L385 (1984).

\bibitem{binder88} K. Binder, D.W. Heermann,
  \textit{Monte-Carlo Simulation in Statistical Physics}, Springer
  (1988).

\bibitem{sokal96} A. Sokal,
  \textit{Monte Carlo methods in statistical mechanics: 
    foundations and new algorithms}, in \textit{Functional
    Integration: Basics and Applications (1996 Carg\`ese Summer School)} 
  ed C. DeWitt-Morette, P. Cartier and A. Folacci 
  (New York: Plenum) pp 131­-192 (1997). \\
  http://www.math.nyu.edu/faculty/goodman/teaching/Monte Carlo/Sokal.ps

\end{thebibliography}
\end{document}